\documentstyle[preprint,aps,epsfig]{revtex}
\tightenlines
\begin{document}

\title{Charmonium production from the hadronic phase}
\author{C. M. Ko$^1$, X. N. Wang$^2$, and X. F. Zhang$^3$}
\address{$^1$Cyclotron Institute and Physics Department, \\
Texas A\&M University, College Station, Texas 77843}
\address{$^2$Nuclear Science Division, \\
Lawrence Berkeley National Laboratory,
University of California, Berkeley, Ca 94720}
\maketitle

\begin{abstract}
Charmonium production from the hadron gas formed in
ultra-relativistic heavy-ion collisions is studied. Using the
$J/\psi$-hadron absorption cross section determined from the
nonperturbative quark-exchange model, which has a peak value
similar to that used in the comover model for $J/\psi$ suppression
and a thermally averaged value consistent from that extracted from
$J/\psi$ data in heavy-ion collisions, we find that $J/\psi$
production from the hadron gas is negligible in heavy-ion
collisions at RHIC energies but is important at LHC energies as a
result of the large number of charm mesons produced at higher
energy collisions. The number of $J/\psi$ produced from these
secondary collisions at LHC may be comparable to that of primary
$J/\psi$'s, which are expected to be dissociated in the quark-gluon
plasma created in the collisions, leading thus to a possible
absence of $J/\psi$ suppression. Similar results are obtained for
$\psi^\prime$ production in ultra-relativistic heavy-ion
collisions.
\end{abstract}

\vspace{0.5cm}
\vspace{0.5cm}
\newpage

One of the signals proposed for identifying the existence of a
quark-gluon plasma in ultra-relativistic heavy-ion collisions is
the suppression of $J/\psi$ production compared to that expected
from the superposition of nucleon-nucleon collisions \cite{matsui}.
According to Matsui and Satz, if a quark-gluon plasma is created in
heavy-ion collisions, $J/\psi$'s produced from initial
nucleon-nucleon interactions will dissociate as a result of Debye
screening and vanishing string tension between $c$ and $\bar c$.
Recent experimental results from heavy-ion collisions at CERN SPS
energies \cite{na38,na50,gonin} have indeed shown a reduction of
$J/\psi$ production. Part of the effects can be attributed to
absorption by nucleons as it is more likely that $J/\psi$ is first
produced as a pre-resonance in a color octet state and thus has a
larger interaction cross section with nucleon
\cite{bodwin,kh,qiao,wong}. Such an effect is needed to account for
the observed $J/\psi$ suppression in proton-nucleus collisions
\cite{gerschel}, where one does not expect the formation of a
quark-gluon plasma. Also, $J/\psi$ absorption by comovers, which
are most pions produced in the collisions, has been suggested
\cite{gavin,capella,gavin1,cassing97}. In collisions of light ions
such as S+U at 200 GeV/nucleon, these absorption mechanisms are
sufficient to explain the experimental data. However, for central
collisions of heavy nuclei such as Pb+Pb at 160 GeV/nucleon, the
anomalous large suppression of $J/\psi$ observed in the experiments
has led to the suggestions that a quark-gluon plasma is formed in
these collisions, and the $J/\psi$ yield is reduced due to
dissociation either by Debye screening \cite{kharz,blazoit,cywong}
or collisions with gluons \cite{shuryak}. If the quark-gluon plasma
has already been formed in heavy-ion collisions at SPS energies, it
is almost certain that it will also be formed in heavy-ion
collisions at the Relativistic Heavy-Ion Collider (RHIC) at the
Brookhaven National Laboratory and the Large Hadron Collider at
CERN (LHC), where energies are much higher. Therefore, $J/\psi$
suppression at RHIC and LHC is considered as one of the most
prominent signals for the quark-gluon plasma.

But will $J/\psi$ production at RHIC and LHC be really reduced?
This depends on whether $J/\psi$ can be regenerated from the
hadronic matter after the phase transition of the quark-gluon
plasma. Such an effect has been shown to be important for hadrons
made of strange quarks \cite{phi}. In hadronic matter, $J/\psi$ can
be produced from the interactions of charm mesons in reactions such
as $D{\bar D}^*, D^*\bar D, D^*{\bar D}^*\to J/\psi\pi$. These
reactions are apparently unimportant at SPS energies but may become
significant at RHIC and LHC since charm production in
nucleon-nucleon interaction increases with center-of-mass energy
while the $J/\psi$ to $c\bar c$ ratio remains essentially at a
constant value of $\sim 2.5\times 10^{-2}$ \cite{gavai}.
Furthermore, the reaction is exothermic as $m_D+m_{\bar D}^*\sim$
3.73 GeV/c$^2$ and $m_{D^*}+m_{{\bar D}^*}\sim$ 3.87 GeV/c$^2$
while $m_{J/\psi}+m_\pi\sim 3.15$ GeV/c$^2$, it is thus more likely
that $J/\psi$'s produced from the hadronic matter will survive and
be detected in experiments. If this is the case, then the final
number of $J/\psi$ may be comparable to or even larger than that of
primary $J/\psi$'s which are dissociated in the quark-gluon plasma,
leading instead to an absence of $J/\psi$ suppression or even an
enhanced $J/\psi$ production. In the following, we shall make an
estimate of $J/\psi$ production from a pion gas formed in heavy-ion
collisions at RHIC and LHC energies.

To model heavy-ion collisions at such high energies, we use the
results from the HIJING calculation \cite{wang}, which takes into
account parton production from semihard scatterings. It shows that
at an initial proper time $\tau_0$ a thermally equilibrated
although chemically non-equilibrated quark-gluon plasma of
temperature $T_0$ is formed. For Au+Au collisions, HIJING predicts
that $\tau_0\sim$ 0.7 and 0.5 fm/c, $T_0\sim$ 0.57 and 0.83 GeV at
RHIC and LHC energies, respectively. The quark-gluon plasma then
cools due to expansion and production of additional partons. It
reaches the critical temperature $T_c\sim 200$ MeV at about
$\tau_c\sim$ 3 fm/c at RHIC and 6 fm/c at LHC, when the quark-gluon
plasma starts to make a transition to a hadron gas, consisting
mostly of pions. Since neither the partons have reached chemical
equilibrium nor the order of phase transition is definitely known,
the time for the system to remain at $T_c$ can not be properly
determined. We shall assume that the proper time at which the
quark-gluon plasma is completely converted to a pion gas is
$\tau_h\sim 2\tau_c$ and will study how the results depend on the
value of $\tau_h$. Since transverse expansion is not expected to be
appreciable in this early phase, the initial volume of the hadronic
matter is simply $V_h\sim\pi R_0^2\tau_h$ if one uses the
boost-invariant model of Bjorken \cite{bjorken}. In the above,
$R_0$ is the radius of the colliding nuclei. Assuming that the
hadron gas expands isentropically, its temperature then decreases
according to the inverse of the cubic root of volume. To include
the transverse expansion of the hadron gas, we introduce an
acceleration $a$. Then its transverse radius increases with time
according to $R(\tau)=R_0+a(\tau-\tau_h)^2/2$ until the velocity
reaches the velocity of light, when the transverse radius increases
linearly with time. From entropy conservation, the temperature of
the hadron gas then decreases as
$T(\tau)=(\tau_h/\tau)^{1/3}(R_0/R(\tau))^{2/3}T_c$. Final hadron
yields and spectra are determined at freeze out with a temperature
$T_f\sim 120$ MeV. We note that for pions in equilibrium the
density is approximately given by $n_\pi\sim 0.25(T/197{\rm
MeV})^3$ fm$^{-3}$.

\begin{figure}
\begin{center}
\centerline{\epsfig{file=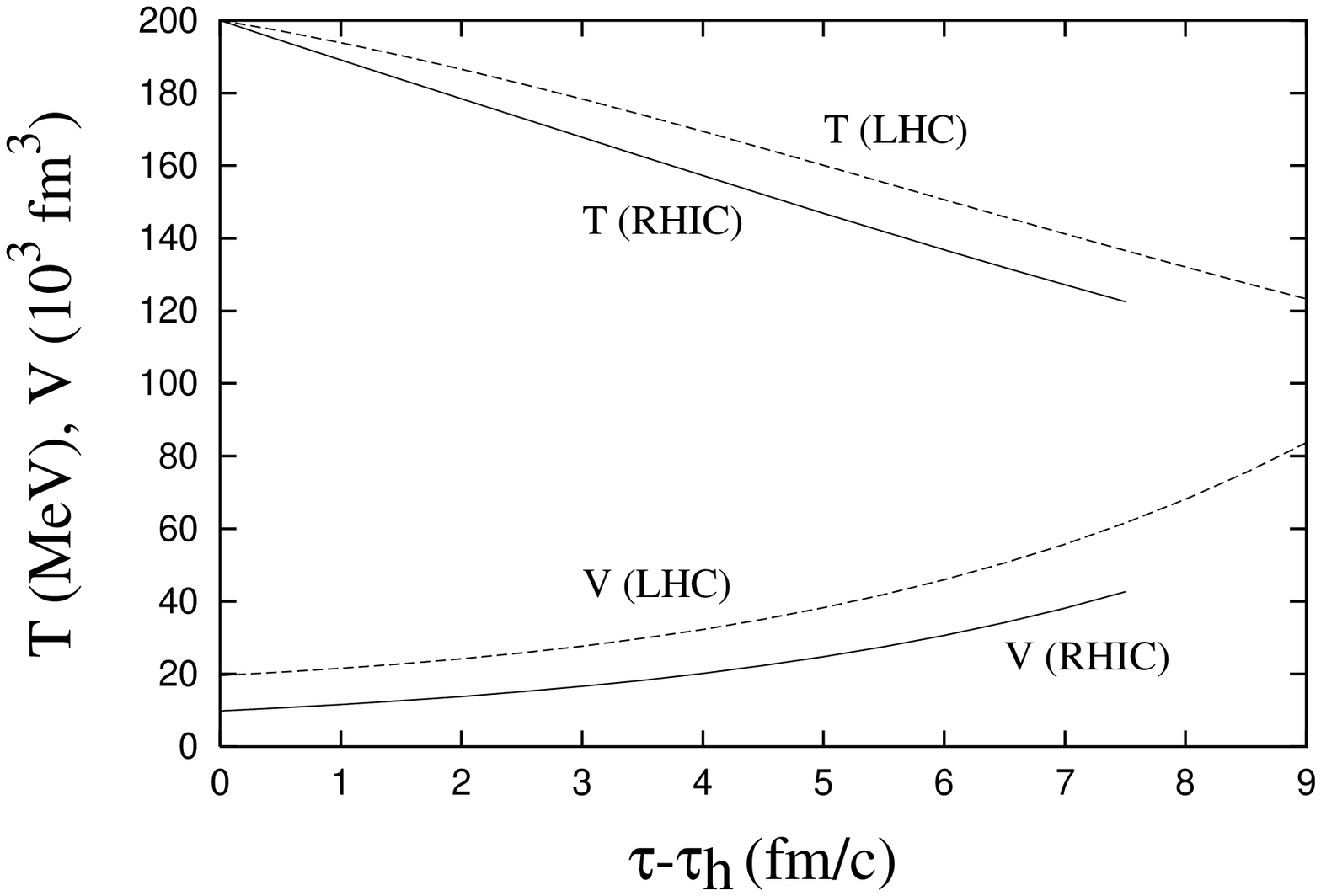,width=5in,height=5in,angle=-90}}
\vspace{1cm}
\caption{Time evolution of the temperature and volume of
hadron gas in Au+Au collisions at RHIC and LHC.}
\label{tv}
\end{center}
\end{figure}

The time evolution of the temperature and volume of the hadron gas
for Au+Au collisions at both RHIC and LHC is shown in Fig. \ref{tv}
for $a=0.1$ c$^2$/fm. The transverse expansion velocity at freeze
out is about 0.75c and 0.9c at RHIC and LHC, respectively. The
initial number of charm quarks produced in these collisions can be
estimated from their production cross section $\sigma_{c\bar
c}^{NN}$ in nucleon-nucleon interaction using \cite{wong94}
\begin{equation}
N_{c\bar c}^{\rm AB}\approx AB\sigma_{c\bar c}^{NN}/\sigma_{\rm in}^{\rm AB},
\end{equation}
where $\sigma_{\rm in}^{\rm AB}$ is the nucleus-nucleus (A+B)
inelastic cross section. For central collisions, $\sigma_{\rm in}^{\rm AB}$
is approximately the nuclear geometrical cross section. The cross
section $\sigma_{c\bar c}^{NN}$ has been evaluated using the
perturbative QCD \cite{wang}, and it has a value 0.16 mb and 5.75
mb for nucleon-nucleon interaction at RHIC and LHC energies,
respectively. For Au+Au collisions, the initial number of $c\bar c$
(i.e., $D\bar D$) pairs, is thus about 3.8 at RHIC and 136 at LHC.
To determine the initial $J/\psi$ number, we assume that the same
$c\bar c$ to $J/\psi$ ratio, about 40, observed in nucleon-nucleon
interaction at center-of-mass energy below 40 GeV \cite{gavai}
holds at RHIC and LHC energies. In this case, the primary $J/\psi$
number is about 0.095 and 3.4 at RHIC and LHC, respectively. Since
a quark-gluon plasma is almost certain to be created in heavy-ion
collisions at RHIC and LHC, both $J/\psi$'s and $D$'s are expected
to be dissociated due to Debye screening. As shown in Ref.
\cite{wang}, thermal charm production from the quark-gluon plasma
is small, it is thus reasonable to assume that the $c$ and $\bar c$
quark numbers remain constant during the quark-gluon phase. After
phase transition to a hadron gas, these charm quarks are more
likely to form $D$ and $\bar D$ mesons and their excited states due
to the large abundance of light quarks in the plasma. Conservation
of charm then requires that the number of charm mesons at the
beginning of the hadron phase is the same as the initial one.

The $J/\psi$ production cross section from $D\bar D$ interaction,
$\sigma_{D\bar D\to J/\psi\pi}$, can be related to the absorption
cross section $\sigma_{J/\psi\pi\to D\bar D}$ via the detailed
balance relation
\begin{equation}
\sigma_{D\bar D\to J/\psi\pi}=d(k_{J/\psi\pi}/k_{D\bar D})^2
\sigma_{J/\psi\pi\to D\bar D},
\end{equation}
where the factor $d$ is 3/4 for $D{\bar D}^*$ and $D^*\bar D$
annihilation, and 1/4 for $D^*{\bar D}^*$ annihilation;
$k_{J/\psi\pi}$ and $k_{D\bar D}$ are, respectively, the relative
momenta of $J/\psi\pi$ and $D\bar D$. The magnitude of $J/\psi$
absorption cross section by pion, $\sigma_{J/\psi\pi\to D\bar D}$,
is still under debate. In the comover model for $J/\psi$
suppression in heavy-ion collisions at SPS energies, it is taken to
be about 3 mb. Both perturbative QCD \cite{kh} and effective
Lagrangian \cite{mat98} calculations give a value which is about a
factor of 10 smaller. On the other hand, studies including
nonperturbative effects via quark-exchange model \cite{martins}
give a peak value of about 6 mb. As shown below, the thermally
averaged effective cross section that takes into account the
threshold effect is much smaller and is consistent with that
extracted in Ref. \cite{capella} from $J/\psi$ suppression data in
heavy-ion collisions. In the following, we shall use the
parameterized values from Ref. \cite{martins}, i.e.,
\begin{equation}
\sigma_{J/\psi\pi\to D{\bar D}}(s)\approx
\sigma_0\left(1-(\frac{s_0}{s})^2\right)\exp[-a(\sqrt{s}-\sqrt{s_1})]
\theta(s-s_0),
\end{equation}
where $s$ is the square of the $J/\psi\pi$ center-of-mass energy and
$s_0$ is the threshold energy. For the final states $D{\bar D}^*$ and
${\bar D}D^*$, the parameters are
$\sigma_0=2.5$ mb, $s_0=15.05$ GeV$^2$, $s_1=17.6$ GeV$^2$,
and $a=11$ GeV$^{-1}$; while for the final state $D^*{\bar D}^*$,
they are $\sigma_0=3.4$ mb, $s_0=16.2$ GeV$^2$, $s_1=19.0$ GeV$^2$,
and $a=11$ GeV$^{-1}$.

\begin{figure}
\begin{center}
\centerline{\epsfig{file=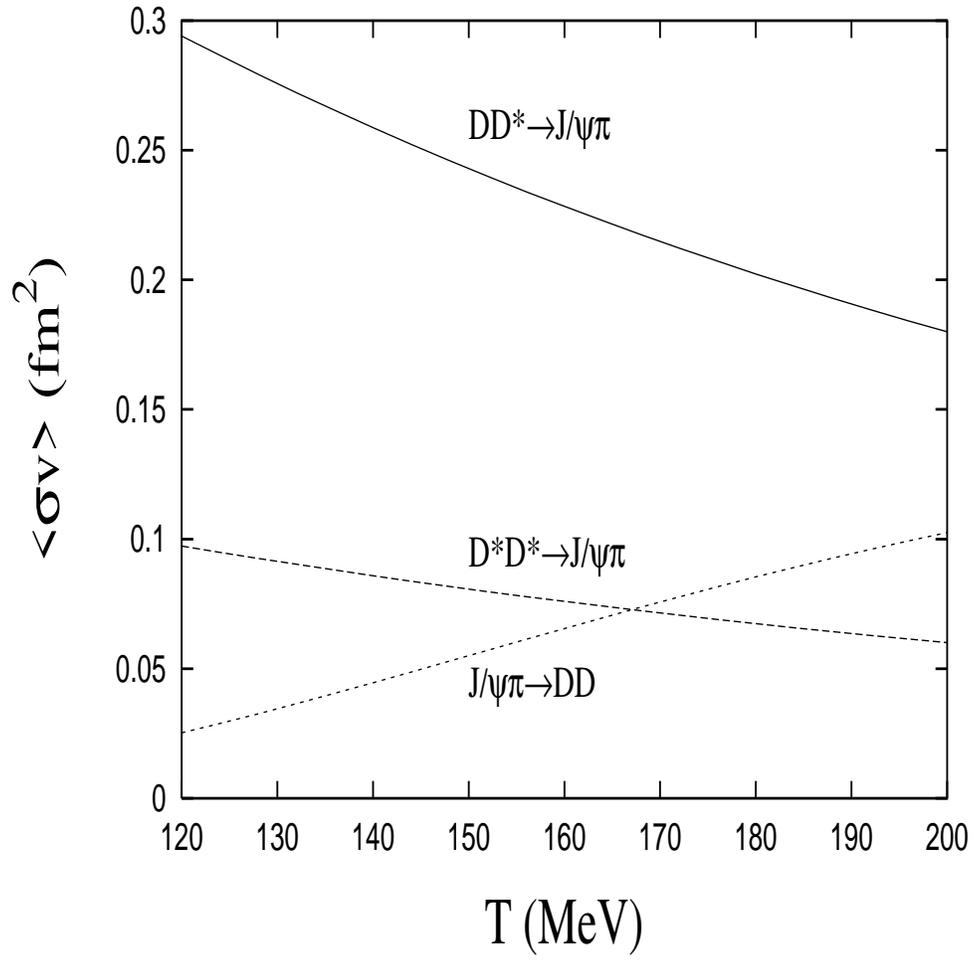,width=5in,height=5in,angle=-90}}
\vspace{1cm}
\caption{Temperature dependence of the thermally averaged cross sections:
$<\sigma_{DD^*}v>$, $<\sigma_{D^*{\bar D}^*}v>$, and
$<\sigma_{J/\psi\pi}v>$.}
\label{sv}
\end{center}
\end{figure}

The quantity needed for estimating $J/\psi$ production from $D\bar
D$ interaction in a hadron gas is the thermal average of
$\sigma_{D\bar D\to J/\psi\pi}v$, where $v$ is the relative
velocity between $D$ and $\bar D$, i.e.,
\begin{eqnarray}
<\sigma_{D\bar D\to J/\psi\pi}v>&=&[4(m_{D_1}/T)^2(m_{D_2}/T)^2
K_2(m_{D_1}/T)]K_2(m_{D_2}/T)]^{-1}
\int_{(m_{D_1}+m_{D_2})/T}^\infty dz\nonumber\\
&\cdot&[z^2-(m_{D_1}/T+m_{D_2}/T)^2][z^2-(m_{D_1}/T-m_{D_2}/T)^2]
\nonumber\\&\cdot&K_1(z)\sigma_{D\bar D\to J/\psi\pi}.
\end{eqnarray}
In the above, $m_{D_1}$ and $m_{D_2}$ are the masses of the two
charm mesons; $K_1$ and $K_2$ are, respectively, the modified
Bessel function of the first and second kind. The above quantity
has been evaluated separately for $D{\bar D}^*$, $D^*\bar D$, and
$D^*{\bar D}^*$ annihilation. In Fig. \ref{sv}, we show the
temperature dependence of $<\sigma_{DD^*}v>$, defined by
\begin{equation}
<\sigma_{DD^*}v>=<\sigma_{D{\bar D}^*\to J/\psi\pi}v>+
<\sigma_{D^*\bar D\to J/\psi\pi}v>,
\end{equation}
and $<\sigma_{D^*{\bar D}^*}v>=<\sigma_{D^*{\bar D}^*\to J/\psi\pi}v>$.
Similarly, one evaluates the
thermal average $<\sigma_{J/\psi\pi\to D\bar D}v>$ for the
production of $D{\bar D}^*$, $D^*\bar D$, and $D^*{\bar D}^*$, and
defines $<\sigma_{J/\psi\pi}v>$ as their sum. The
temperature dependence of $<\sigma_{J/\psi\pi}v>$ is also shown in
Fig. \ref{sv}. We see that $<\sigma_{D^*{\bar D}^*}v>$ is about
a factor of three smaller than $<\sigma_{D\bar D}v>$ due to
a smaller cross section. Furthermore, $<\sigma_{D\bar D}v>$ is larger than
$<\sigma_{J/\psi\pi}v>$ at all temperatures. The value of latter
ranges between 0.25 to 1.0 mb, which is much smaller than the cross
section itself as a result of threshold effects so only pions with
sufficient energy can destroy a $J/\psi$.

The rate of $J/\psi$ production from the hadron gas is given by
\begin{equation}\label{rate}
\frac{dR}{d\tau}=<\sigma_{DD^*}v>n_Dn_{D^*}
+<\sigma_{D^*{\bar D}^*}>n_{D^*}^2
-<\sigma_{J/\psi\pi}v>n_{J/\psi}n_\pi,
\end{equation}
where $n_D$ and $n_{D^*}$ are the densities of $D$ and $D^*$,
respectively, and their relative value is determined by assuming
that they are in chemical equilibrium, i.e.,
\begin{equation}
\frac{n_D}{n_{D^*}}=\frac{1+(1/3)\exp((m_{D^*}-m_D)/T)}
{1+3\exp(-(m_{D^*}-m_D)/T)}.
\end{equation}
In Eq. (\ref{rate}), we
have used the fact that the densities of charm and anti-charm
mesons are the same. The density of $J/\psi$ is denoted by
$n_{J/\psi}$.

\begin{figure}
\begin{center}
\centerline{\epsfig{file=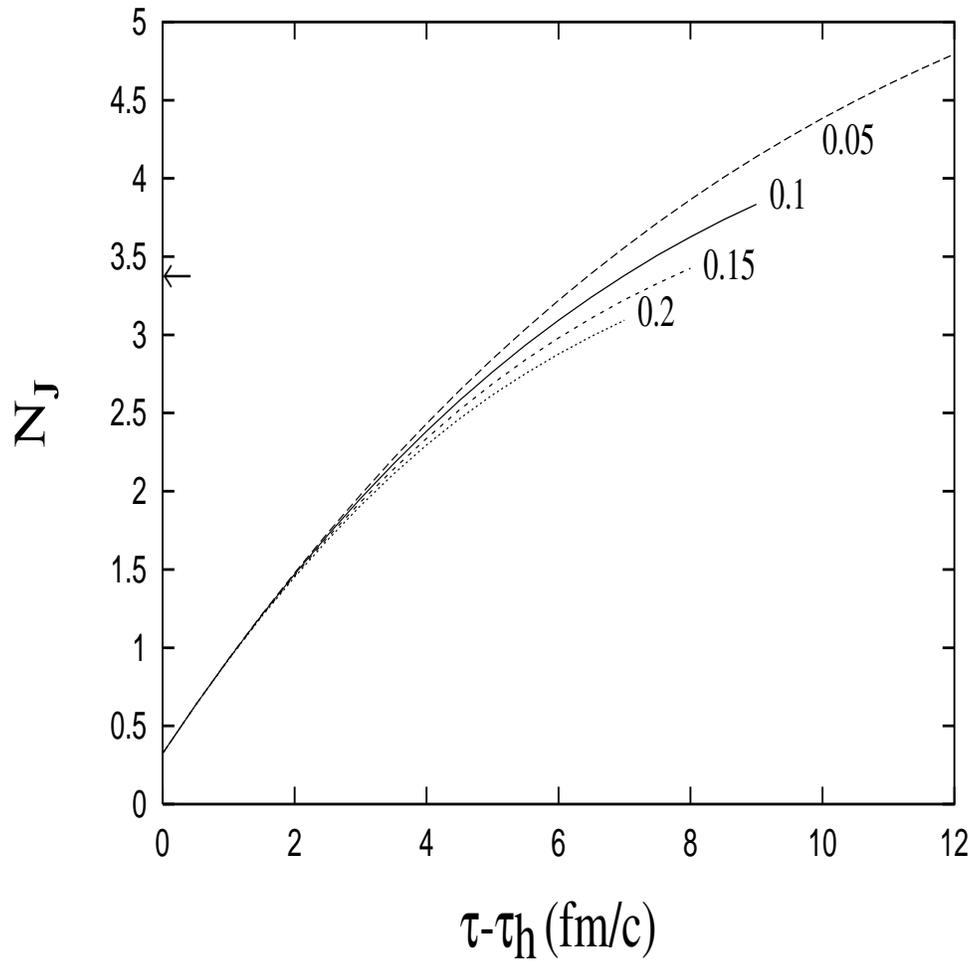,width=5in,height=5in,angle=-90}}
\vspace{1cm}
\caption{Time evolution of the abundance of $J/\psi$ and $D$ from Au+Au
collisions at LHC for different values of transverse acceleration
$a$ (c$^2$/fm). The initial $J/\psi$ number is denoted by the left arrow.}
\label{num}
\end{center}
\end{figure}

The total number of produced $J/\psi$ is obtained by multiplying
the above rate by the volume of the hadron gas and then integrating
over time. We find that at RHIC the number of $J/\psi$ produced
from $D\bar D$ annihilation is only 0.005 and is more than an
order-of-magnitude smaller than the number of primary $J/\psi$.
This is different at LHC as a result of the much larger number of
$D\bar D$ present in the hadron gas. In Fig. \ref{num}, we show by
the long-dashed curve the $J/\psi$ number calculated with a
transverse acceleration $a=0.1$ c$^2$/fm as a function of time in
Au+Au collisions at LHC. It is seen that the $J/\psi$ number
increases with time and at freeze out is comparable to the number
of primary $J/\psi$, which is indicated by the left arrow. We note
that calculations carried out for heavy-ion collisions at SPS
energies also show a negligible effect on $J/\psi$ production from
$D\bar D$ annihilation in the hadron gas.

For larger values of $a$, the hadron gas expands faster so its
temperature drops more appreciably, leading to a reduction of the
number of produced $J/\psi$ as shown in Fig. \ref{num} for $a=0.15$
and 0.20 c$^2$/fm. However, the transverse expansion velocity in
these cases reaches the velocity of light after a few fm/c, which
seems unreasonable. On the other hand, with smaller values of $a$,
the final $J/\psi$ number is increased as shown in Fig. \ref{num}
for $a=0.05$ c$^2$/fm.

If pions are out of chemical equilibrium \cite{song}, $J/\psi$
absorption may be enhanced. Using a pion chemical potential of 130
MeV, the final $J/\psi$ number is reduced only slightly and is
still comparable to the primary one. The lifetime of the phase
transition also affects the result. If it is doubled, then the
number of produced $J/\psi$ is reduced to about 2.25, while it is
increased to 6.15 if the phase transition is instantaneous.
Reducing the value of the temperature at which the phase transition
occurs does not change much the number of $J/\psi$ produced from
the hadron gas. Using $T_c=150$ MeV, the $J/\psi$ number is reduced
only slightly to 3.53 due to the competing effects of increased
$<\sigma_{DD^*}>$ and shortened lifetime of the hadron gas with
decreasing temperature.

The dependence of $J/\psi$ production on the cross section
$\sigma_{J/\psi\pi\to D\bar D}$ is more significant. If the cross
section is an order of magnitude smaller than the one used here as
predicted by the perturbative QCD \cite{kh} and the effective
Lagrangian \cite{mat98} calculation, then the number of $J/\psi$
produced from the hadron gas at LHC is only 0.41 and is much less
than that of primary ones. It is thus important to have a better
determination of this cross section from both experiments and
theoretical models.

The above analysis can be generalized to $\psi^\prime$ production
in ultra-relativistic heavy-ion collisions. The primary
$\psi^\prime$ number is about a factor of 5 smaller than the $\psi$
number, i.e., about 0.019 and 0.682 at RHIC and LHC, respectively.
Since the radius of $\psi^\prime$ is about twice that of $\psi$,
its absorption cross section by a pion is about four times larger.
Using this cross section and the $\psi^\prime$ mass, we obtain only
0.002 $\psi^\prime$ from $D\bar D$ annihilation in hadron gas at
RHIC, again negligible as in the case of $J/\psi$. For heavy-ion
collisions at LHC, the number of $\psi^\prime$ produced from the
hadron gas is about 1.15 and is more than the number of primary
$\psi^\prime$. Therefore, there will not be a suppression of
$\psi^\prime$ production at LHC either.

To summarize, we have estimated the number of $J/\psi$ produced
from the reaction $D\bar D\to J/\psi\pi$ in the hadron gas formed
in heavy-ion collisions at RHIC and LHC using the initial
conditions determined from the HIJING parton model. Although this
is a negligible effect at RHIC, it may become important at LHC as a
result of the appreciable number of $D$ and $\bar D$ mesons in the
hadron gas. We have found that the number of $J/\psi$ produced from
this reaction at LHC is comparable to that produced from initial
primary collisions, which are either absorbed by nucleons or
dissociated in the quark-gluon plasma formed in the collisions. We
thus conclude that in heavy-ion collisions at LHC one may not see a
suppression of $J/\psi$ production. A similar estimate has been
made for $\psi^\prime$ production, where the number of
$\psi^\prime$ produced from the hadron gas may be even larger than
the primary ones. To use the $J/\psi$ and $\psi^\prime$ yields as
signals for the quark-gluon plasma at LHC therefore requires a good
understanding of their production from the hadron gas.

\bigskip

This work was started while the authors were visiting the Institute
for Nuclear Theory at University of Washington for the program on
Probes of Dense Matter in Ultrarelativistic Heavy Ion Collisions,
and they thank Y. Asakawa, C. Gale, J. Kapusta, V. Koch, and C. Y.
Wong for useful discussions. The work of CMK and XFZ was supported
in part by the National Science Foundation under Grant No.
PHY-9509266 and PHY-9870038, the Welch Foundation under Grant No.
A-1358, and the Texas Advanced Research Program. The work of XNW
was supported by the Director, Office of Energy Research, Division
of Nuclear Physics of the Office of High energy and Nuclear Physics
of the U.S. Department of energy under Contract No.
DE-AC03-76SF00098.

\newpage

\newpage

\newpage

\end{document}